# Optoelectronic Readout of single Er Adatom's Electronic States Adsorbed on the Si(100) Surface at Low Temperature (9K)

Eric Duverger[1] and Damien Riedel[*2]

[1]*Institut FEMTO-ST, Univ. Franche-Comté, CNRS, F-25030 Besançon, France.*

[2]*Institut des Sciences Moléculaires d'Orsay (ISMO), CNRS, Univ. Paris Sud, Université Paris-Saclay, F-91405 Orsay, France.*

*Abstract :*

**Integrating nanoscale opto-electronic functions is vital for applications such as optical emitters, detectors, and quantum information. Lanthanide atoms show great potential in this endeavor due to their intrinsic transitions. Here, we investigate Er adatoms on Si(100)-2x1 at 9K using a scanning tunneling microscope (STM) coupled to a tunable laser. Er adatoms display two main adsorption configurations that are optically excited between 800 nm and 1200 nm while the STM reads the resulting photocurrents. Our spectroscopic method reveals that various photocurrent signals stem from the bare silicon surface or Er adatoms. Additional photocurrent peaks appear as the signature of the Er adatoms relaxation, triggering efficient dissociation of nearby trapped excitons. Calculations using the density functional theory with spin-orbit coupling correction highlight the origin of the observed photocurrent peaks as specific 4f→4f or 4f→5d transitions. This spectroscopic technique can pave the way to an optoelectronic analysis of atomic and molecular assemblies by offering unique insight into their intrinsic quantum properties.**

* Corresponding author

**Main**

Addressing electronically and optically atomic scale devices such as in a molecular assemblies or single atoms on surfaces is a tremendous challenge that has constantly fueled both fundamental and applied research over recent decades[1,2,3,4]. Significant progress in this domain has focused on diverse aspects, among which, mapping the optical electromagnetic field in the junction of a scanning tunneling microscope (STM)[5,6], Raman spectroscopy at the molecular scale[7] or probing surface carrier dynamics with pump-probe excitation[8,9]. Conducting this kind of investigation remains challenging. One major hurdle lies in effectively coupling laser light to the STM junction, ensuring the STM operates normally (measuring tunnel current, bias voltage, and tip surface distance) while maintaining control over the electromagnetic field parameters beneath the STM tip apex, including polarization, intensity, and spatial distribution[10]. Proper coupling of the laser light with the tip-surface area as aforementioned, allows extraction of specific information from the surface and/or adsorbates[11,12]. An instance of this type of control occurs when the laser beam interacts with the STM junction area to acquire precise vibrational data, once directed outside the STM chamber[13,14]. Another approach involves the analysis of photoelectron signals from optically excited adsorbates, as for example, when HOMO-LUMO frontier orbitals transitions occur[15]. One fascinating aspect of this last approach that, to date, has never been realized, is to describe the precise electronic structure of a single atom adsorbed on a surface by reading locally spectroscopic photocurrent information related to its electronic state. This concept gains relevance by investigating lanthanide (Ln) atoms that offer long coherence times finding interests for quantum applications[16]. Yet, successful investigations were able to probe various buried atoms, including Ln ions, embedded in polycrystalline silicon with the use of a complex mesoscale framework, revealing insight from the collective aspect of the implanted ions in the device[16,17,18].

Our study focuses on analyzing the induced photocurrent resulting from optically excited Er adatoms adsorbed on the bare Si(100)-2x1 surface at low temperature (9 K) to provide a specific and precise spectrum of its excited electronic structure. Each of the two observed Er adsorption configurations is optically addressed by precisely positioning the STM tip apex over the Er atom and directing a pulsed laser train at the STM junction (Fig. 1a). At negative bias, the photoinduced tunnel current measured by the STM in response to laser irradiation consists of two primary components: (i) The first one is arising from the bare silicon surface when optically excited. (ii) The second contribution originates from the optical excitations of the Er adatom itself. Detailed numerical simulations using the density functional theory (DFT) with a spin orbit coupling correction allow describing the excited electronic structure of each Er adatom configuration. Our combined experimental and numerical simulations allow distinguishing between the two adsorption sites and gain insight into the mechanism behind the induced photocurrent processes. Our findings indicate that the observed photocurrent peaks are the signature of an efficient dissociation of trapped excitons located within the probed Er adatoms. Their dissociation is triggered by the relaxation of the optically excited erbium ($Er^*$) adatom via various suggested processes[11].

**Er adatom adsorption sites and electronic configuration**

The most common Er adsorption sites on Si(100) surface are of two main types (Fig. 1b). One is atop silicon dimer row, in pedestal configuration, erbium top ($Er^T$). The second site is located at the back-bond row of Si(100)-2x1 structure, named erbium back-bound ($Er^{BB}$). These sites resemble adsorption of 3d-5d transition metals on semiconductor surfaces[19]. More details about these sites are given in Figs. 1c, 1d with ball-and-stick structures obtained from numerical relaxation calculations of the bare silicon surface with an Er adatom and As dopant. Changing surface bias from negative to positive values allows imaging and comparing occupied/unoccupied STM topographies of $Er^{BB}$ and $Er^T$ with DFT simulations (Figs. 1e and 1f). The $Er^{BB}$ restatom (Fig. 1e) lies at the back bond row of Si(100)-2x1, positioned between two silicon dimers. The $Er^T$ adatom is straddling on top of a single silicon dimer in a pedestal configuration (Fig. 1f). The STM images of Er adsorption sites are accurately reproduced (lower panel in Figs. 1e, 1f), for both biases, via DFT simulations using slabs similar to Figs. 1c and 1d (refer to Fig. S16 for more details). In particular, the dark halo surrounding the $Er^{BB}$ is replicated at positive bias, as well as the spatial brightness distribution of the $Er^T$ adatom for all biases. One can notice that the spatial distribution of the density of state (DOS) observed in the STM images of the Er adatoms exhibits rather elliptical (Fig. 1f) or spherical (fig. 1e) shapes. These distinctive outlines may be the signature of the spatial distribution of the occupied Er:6s,5d orbitals which inherently possess a larger spatial extension[20]. In this situation, the direct observation of the Er:4f orbitals, which are spatially very localized and usually screened by the Er:5d orbitals is scarcely achievable[21,22].

To characterize the electronic structure of the Er adatoms without optical excitation, we measure the dI/dV spectroscopy curves with the STM, which illustrates the variation of DOS of the adsorbates as a function of the bias voltage. These measurements are depicted in Figs. 1g and 1h when acquired on top of the $Er^{BB}$ and $Er^T$, respectively (red curves), and compared with the dI/dV spectroscopy obtained on the bare Si surface (blue curve). Both dI/dV spectra acquired on Er adatoms are mainly composed of three intense DOS peaks in the occupied states and a continuum of DOS in the unoccupied states. It is important to highlight that the energy of each dI/dV peak exhibits slight shifts between the two Er sites, and their intensities undergo changes, suggesting distinctly noticeable interactions between the surface and the two adatom configurations. This difference is further validated by comparing these data with the dI/dV spectrum collected on the bare Si(100) surface.

The DOS peak energies observed at the $Er^{BB}$ site (Fig. 1g) manifest two prominent peaks of high intensities at -0.9 V and -1.5 V, lying within the energies of the $Si_\sigma$ (-1.4 eV) and $Si_\pi$ (-0.9 eV) bands. This alignment reflects possible interactions with the four surrounding Si atoms when adsorbed within the silicon back-bond row. Conversely, at the $Er^T$ site (Fig. 1h), the DOS peak energies exhibit significant shifts (-1.1 V and -1.6 V) compared to the bare silicon surface $Si_\pi$ and $Si_\sigma$ band states. It suggests weak chemical bonding between the adsorbates and the silicon surface dimers atoms that usually involve $Si(sp^3)$ orbitals.

It is important to mention that the Si(100)-2x1 surface is known to be very reactive, and that the high concentration of As dopant in our substrates induces the formation of alternatively buckled silicon dimer rows[23,24] following a charge distribution changing from positive ($Si_{up}$) to negatively charge ($Si_{down}$) states. Such configuration may significantly impact the bonding of Er adatoms at the Si(100) surface, differently to what occurs when Er atoms are located deeper in the bulk silicon[16]. Notably, it is often considered that the 4f orbitals of the Er atom in the silicon bulk have only weak interactions with the valence electrons of the hosting crystal, or that these interactions are neglected[25].

Further Bader charge analysis of the $Er^T$ and $Er^{BB}$ electronic structure shows that their charge state is slightly positive and worth + 0.564 ē and + 0.839 ē, respectively (Fig. S5 and note 1). For both Er adatom configurations, the Er:6p orbitals exhibit only weak interaction with the surface while the Er:6s orbital hybridizes with the Si:3p orbitals. This hybridization leads to avoid the Er:5d and Er:4f orbitals from substantial electronic interactions with the surface. However, it has also been established that Er:4f orbitals can hybridize with silicon despite their very localized nature[26,27]. Therefore, in their adsorbed configuration, the charged state of the Er adatoms can be denoted $Er^T$:[Xe] $6s^1 5d^{0.9} 4f^{11.4}$ and $Er^{BB}$:[Xe] $6s^{0.95} 5d^{0.8} 4f^{11.5}$. Here, the Er:4f orbitals are slightly depopulated to partially fill the Er:5d orbitals to gain stability of the structure[28]. Interestingly, unlike the previously observed ground state of embedded $Er^{3+}$:[Xe]$4f^{11} 6s^0 5d^0 (^4I_{15/2})$ atoms in bulk silicon, our simulations show that the ground states of the two adatoms configurations are $Er^T$:[Xe]$6s^1 5d^{0.9}(^1S_0) 4f^{11.4}(^3D_3)$ and $Er^{BB}$:[Xe]$6s^{0.95} 5d^{0.8}(^1S_0) 4f^{11.5}(^3S_1)$. To keep a most accurate description of their electronic structure, the above description of the Er adatoms is given with partial charge states (see supplementary documents Figs. S1 to S5 for additional details).

**Photocurrent measurements at the bare silicon surface.**

When the STM tip is located far from any Er adatoms during the laser irradiation, specifically over the bare silicon surface, we can detect a characteristic photocurrent signal by tuning the incoming laser wavelength (Fig. 2a). Under these conditions, the ensuing photocurrent contribution consists of a large photoelectron band emission lying from 800 nm to 1200 nm (black curve in Fig. 2a). Within this broad photocurrent signal, one can observe two sharp photocurrent peaks: one is located at 1030 nm and the second one is measured at 1064 nm. One can compare the photocurrent emission signal in Fig. 2a with the intensity of the incoming laser measured at the output of the optical fiber of the laser (red curve in Fig. 2a). The average laser power varies between 5.6 mW at 800 nm to 3.4 mW at 1200 nm (see Figs S8 and S10 for details). The other photocurrent peak measured at 1064 nm (1.16 eV) can be distinctly ascribed to the very intense contribution of the incident laser light (internal pump laser), which is up to 3 times more powerful than the rest of the laser spectrum (~ 13 mW). The second photocurrent peak observed at 1030 nm (1.2 eV) is not related to any laser intensity peak and lies at an energy being very close to the band gap energy of the bare silicon surface. Experimentally, the surface gap energy of the Si(100)-2x1 can be estimated to be in the range of 1.0 eV to 1.2 eV when measured with dI/dV spectroscopy[24]. The indirect bulk silicon gap energy is rather well documented and worth 1.12 eV[29] leading to an energy difference with the observed photocurrent peak at 1030 nm of ~ 80 meV. It is relevant to notice that the photocurrent signal as observed in Fig. 2a is very similar for positive and negative biases (see Fig. S10).

To further understand the photocurrent spectrum observed on the bare Si(100) surface at positive bias, we have acquired a set of photocurrent curves for various decreasing tip surface distances dz-$\delta z$ where $\delta z$ is varying between $\delta z = 0$ Å to a maximum value of $\delta z = 2$ Å (Fig. 2b). From the curves presented in figure 2b, we have extracted the variation of the photocurrent values, for increasing $\delta z$ values. Fig. 2c gathers these data for three chosen wavelengths fixed at 910 nm, 1030 nm and 1064 nm.

The three sets of values obtained at 910 nm, 1030 nm and 1064 nm shown in Fig. 2c can be numerically fitted by an exponential term (red curves as $I(\delta z) = I_0 + I_a \exp(\kappa\, \delta z)$) traducing the known dependence of the tunnel current in an STM junction as a function of the tip-surface variations. This phenomenon confirms that the induced photocurrent, at these wavelengths, arises from genuine tunnel electrons. However, while the ensuing κ values worth 1.68 Å$^{-1}$ and 2.04 Å$^{-1}$ at 910 nm and 1030 nm, κ =0.68 Å$^{-1}$ at 1064 nm, indicating that the tunnel transport is more perturbed at this wavelength[10].

To delve the properties of the detected photocurrent arising from the irradiated bare silicon surface, we have plotted the variation of the photocurrent intensity at a static tip surface distance ($\delta z$ =1.4 Å) when the surface bias varies from -2.0 V to +2.0 V for the three considered excitation wavelengths (Fig. 2d). These measurements indicate that the photocurrent slowly increases at positive biases up to 2V. At negative biases, the photocurrent variations exhibit a sharp decrease at a voltage threshold ~ -1.2 V. Because of the presence of the silicon surface energy band gap, photocurrent measurements cannot be performed near zero bias voltage. Connecting the boundary values between positive and negative biases data via straight lines (Fig. 2d) enables to estimate the photocurrent at zero bias to be within zero for irradiations at 910 nm and 1030 nm while slightly positive at 1064 nm. In line with this, the variation in photocurrent induced at 1064 nm, as depicted in Fig. 2d, seems to display less dependency on bias voltage. Furthermore, the current drop at -1.2 V is not clearly observed for this wavelength. As the excitation wavelength is tuned from 800 nm (~ 1.55 eV) to 910 nm (~ 1.36 eV), the observed photocurrent spectrum arising from the bare silicon surface is therefore composed of an envelope signal on which two photocurrents peaks at 1030 nm and 1064 nm are added. Considering the photocurrent curves contribution for various surface bias voltages (Fig. 2d), we have chosen, in the following, to compare the photocurrent arising from the Er adatoms for a fixed voltage of -1.2 V. The origin of the photocurrent peak observed at 1030 nm will be discussed in detail in the last section of this article[30].

**Contribution of the photocurrent peaks measured at the Er adatoms**

We optically probe each of the two distinct Er adatom configurations by placing the STM tip apex above it while the STM junction is illuminated with a pulsed laser train. For an optical excitation varying from 800 nm to 1200 nm, the Er photocurrent spectra acquired at the Er$^{BB}$ and Er$^T$ adatoms appear to be composed of the previously observed photocurrent contribution on top of which are superposed very narrow photocurrent spikes as observed in Figs. 3a and 3b. The two photocurrent peaks previously observed on the bare Si surface at 1030 nm and 1064 nm are consistently identified in addition to other sharp photocurrent peaks of various intensities. For the Er$^{BB}$, the major photocurrent peaks (Fig. 3a) are discerned at 898 nm, 925 nm, 933 nm and 1049 nm. Similarly, the main photocurrent peaks observed at the Er$^T$ are more profuse and observed at 886 nm, 918 nm and 959 nm with two closely narrow spaced peaks at 967 nm and 970 nm, as well as two other peaks at 1017 nm and 1121 nm. Other examples of photocurrent spectrum acquired on several Er adatoms are provided in Fig. S18.

To elucidate the origin of the photocurrent peaks retrieved from the Er adatoms, we have performed a comprehensive study of the electronic structure of each Er adatom by extracting the equivalent Kohn-Sham energies localized at each maximum of the Er:4f and Er:5d orbitals from the calculated projected density of states (PDOS) curves. The major trends of this analysis are presented in Figs. 3c and 3d for the spin-up and spin-down components of the Er:5d and Er:4f orbitals of the Er$^{BB}$ and Er$^T$, respectively. One can have a look at the PDOS curve in Figs S1 to S4 for further details.

Despite some differences, the calculated electronic structure for both Er adatoms shows some similar features: a few occupied 5d orbitals components lie below the surface Fermi level energy while most of the unoccupied states of the 5d orbitals span various energy ranges for both spins. In regards to the Er$^{BB}$:4f orbitals, they are mostly occupied and the empty Er$^{BB}$:4f states are restricted to a few components. The spin-up Er$^T$:4f orbitals are all occupied while the spin-down Er$^T$:4f orbitals mostly span around the silicon band gap energy with also a few empty states components. The electronic configurations of these Er adatoms favor various Er:4f $\rightarrow$ 5d and Er:4f $\rightarrow$ 4f transitions. Such electronic structures lead to a spin magnetic moment of ~ 2.4 $\mu_B$ for Er$^{BB}$ and 2.5 $\mu_B$ for Er$^T$, ($\mu_B$ being the Bohr magneton), mostly carried by the 4f orbitals of both adatoms types (see Fig. S5). Taking into account the electronic configuration that involves hybridization between the Er:4f and Er:5d orbitals[26], it is manifest that the electronic interactions between the Er:4f orbitals and the Si surface remain relatively weak although some Er:4f orbitals may appear as almost pinned near the silicon Fermi level energy (e.g. Er$^{BB}$:4f$_{xz^2}$).

Based on this analysis, it is possible to estimate which electronic transition could mainly participate to the optical excitation of the Er adatoms and thus to the photocurrent. In this investigation, we will consider any type of spin transition since Laporte parity rule is often violated in solid-state systems. In our experimental conditions, specific optical transitions may also be allowed when associated with vibrational excitation modes of the surface[31]. Some of these vibrations can be induced by the tunnel electron flux via inelastic transitions and might also be associated with the optical excitation of the indirect band gap of the Si(100) bulk[32].

From the Khon-Sham energies indicated in Figs. 3c and 3d, we can select the most probable optical transitions that fit the wavelengths at which the photocurrent peaks appear on each Er adatoms. These transitions are described in table 1.

For the Er$^{BB}$ adatoms (listed in Table 1), three experimental photocurrent peaks can primarily be ascribed to Er:4f → 4f orbitals transitions, with the exception of the photocurrent peak at 898 nm, for which the calculated transition 4f$_{↓yz^2}$ → 5↑d$_{z^2}$ lying at 890.42 nm is the most accurate. The three other photocurrent peaks observed experimentally on the Er$^{BB}$ involve the 4f$_{↑z3}$ → 4f$_{↓xz^2}$, the 4f$_{↓z(x^2-y^2)}$ → 4f$_{↓xz^2}$ and the 4f$_{↓yz^2}$ → 4f$_{↓xyz}$ transitions. For the Er$^{BB}$:4f → 5d transition, the variation of quantum number are |Δl| = 1 with |Δs| =1 whereas the variation of total angular momentum and magnetic quantum number are |ΔJ| =3/2 and |Δml| = 1, respectively. This may lead to a lower transition probability at this wavelength compared to the 3 other ones. For the 4f → 4f transitions in Er$^{BB}$, the variation of total angular momentum |ΔJ| =1 while the variation of magnetic quantum number |Δm$_l$| remains within unity, which indicates a high level of transition probability. Among these three transitions, only one shows a spin flip (|Δs| =1) at 925 nm within the same 4f orbital, indicating that vibrational modes of the Er-Si are probably involved[33]. From the calculated electronic state on the Er$^{BB}$ adatoms using SOC correction, we observe that the averaged calculated secondary total angular momentum m$_j$ =1.27 ± 0.04 and |Δm$_j$| = 0. When this information is combined with a variation of |Δs| =1, the ensuing transition may indicate an intermediate coupling (jj coupling) in addition to the L.S one[34]. For the Er$^{BB}$ adatom, the averaged energy shift between the calculated transitions and the experimental ones is |Δν| =11.85 ± 5.6 cm$^{-1}$ (excluding the 4f → 5d), indicating the high accuracy of our DFT simulations and the reliability of our method for estimating excitation transitions.

The experimental photocurrent peak assignments for the Er$^T$ site are determined using similar method involving the minimization of the energy shift between the calculated and the experimental transition energy values. For the Er$^T$ adatom, the average energy shift between the calculated transitions and the experimental ones amounting to |Δν| =50.7 ± 22.4 cm$^{-1}$. The second half of Table 1 reveals six 4f → 4f and one 4f → 5d transitions for the Er$^T$ adatom having a high level of probability. All the calculated 4f → 4f transitions involve |Δm$_l$| =0, or ± 1 while |Δs| = 0. The calculated 4f → 5d transition at 956.32 nm indicates a variation of total angular momentum |ΔJ| =3/2. Due to its relatively high |ΔJ| value, such transition may have a lower probability of occurrence, compared to the other ones, as

observed experimentally. For $Er^T$ adatom, the averaged calculated secondary total angular momentum $<m_j>=0.32 \pm 0.06$ and its variation worth $|\Delta m_j| = 0$, indicating that most of the selected 4f → 4f transitions have a relatively high probability to occur. It is particularly the case for the two transitions that are regularly observed at 967 nm and 970 nm for various $Er^T$ adatoms (see Fig. S18).

We have been able to assign the various photocurrent peaks detected in our experiments to the various Er excited states with a relatively good precision. Considering that the pulsed laser used for this work has a bandwidth of 2 nm (i.e. $\Delta \nu$ ~ 17 cm$^{-1}$ at 1100 nm and 31 cm$^{-1}$ at 800 nm), it is expected that the accuracy in the comparison between the photocurrent peak energies and the calculated optically excited states of the Er adatoms will reach similar spectral limits. However, it is interesting to understand the possible origins of this energy shift. One parameter that may induce an energy shift of the Er electronic levels is the electrostatic field applied at the STM junction via the bias voltage. Considering that the electrostatic field **E** under the STM tip can easily reach values up to **E** ~ $10^{10}$ V.m$^{-1}$ (i.e. **E** = 0.2 V.Å$^{-1}$) in our experimental conditions, it is relevant to examine if the estimated energy shift is correlated to the influence of **E** on the Er:4f and Er:5d orbitals energies of the Er adatoms. To confirm this hypothesis, we have performed additional DFT calculations of the electronic structure of each Er adatom when an electrostatic field is applied in a direction perpendicular to the surface. The wavenumbers shifts are computed for each of the Er:5d and Er:4f orbitals energies peaks arising from the energy difference with and without electrostatic field (Fig. S14).

For these calculations, we can notice that the Er:4f orbitals energies are all shifted by the same small amount of energies $<\Delta \nu>$ ~ 55 ± 4 cm$^{-1}$ for **E** = 0.3 V.Å$^{-1}$. Thus, the Er:4f → 4f optical transitions wavelength will not be visibly impacted by the presence of the electrostatic field. However, the Er:5d orbital energies peaks are differently altered when the electrostatic field increases (Fig. S14). This calculation reveals that the variations of the $Er^{BB}$:5d orbitals energies are in the range 0 to 2000 cm$^{-1}$ whereas it varies between 0 and 600 cm$^{-1}$ for $Er^T$:5d orbitals for **E** varying between 0 and 0.4 V/Å. These values are significantly larger than the averaged energy shift observed in Table 1 (i.e. 11.85 cm$^{-1}$ for $Er^{BB}$ and 50.7 cm$^{-1}$ for $Er^T$) since most of the photocurrent peaks observed experimentally are assigned to Er:4f → 4f transitions. Additionally, no substantial changes in wavelength are observed in the photocurrent peaks across the range of measured spectra. This provides strong evidence that the energies at which the photoelectron peaks are detected remain unaffected by the electrostatic field present at the STM junction.

**Discussion about the physical processes involved in the photocurrent.**

During the irradiation of the STM junction with the laser, when the STM tip is located on top of the bare silicon surface, several types of processes may occur (Fig. 4). When the laser wavelength is tuned from 800 nm to 1200 nm, the irradiated zone that surrounds the STM junction promotes the formation of a large number of free charge carriers in the silicon surface as well as hot electrons in the STM tip. Hence, free carriers and free excitons are simultaneously created at various energies, ranging from 1.55 eV (800 nm) to 1.03 eV (1200 nm) within the Si(100) conduction band, specifically from the top of the $Si_{\pi*}$ orbital band to the lower edge of the conduction band. Meanwhile, the electronic population at the tungsten tip at the Fermi level energy is modulated via the formation of hot electrons[35]. This transient electronic population change may significantly affect the charge state of the STM junction and hence the shape of the tunnel barrier[12].

At positive bias (Fig. 4a), when the STM tip probes the bare silicon surface, the main source of tunnel electrons arises from photocreated hot electrons in the tungsten tip while most of the free charges created in the silicon surface will be rapidly evacuated towards the mesoscopic contacts of the sample. The ensuing tunnel electrons can thus probe the shape of the empty states $Si_{\sigma*}$ band. Additional processes involving the neutralization of positive charges in the valence band (not shown) or near the edge of the conduction band may also occur as well as the dissociations of free excitons created in the Si(100) surface during the interaction with the incoming tunnel electrons from the STM tip[11]. Tunnel electrons can also travel back to the STM tip, processes that are indeed made possible when the initial electronic population of both sides of the junction are perturbed by the laser field rendering observable the photocurrent peak at 1030 nm and 1064 nm that arises from the photoexcited Si(100) surface[12].

When a negative bias voltage is applied at the STM junction while the STM tip probes the bare silicon surface (Fig. 4b), photo-created free carriers in the Si(100) conduction band will prominently contribute to the photocurrent signal resulting in a photocurrent envelope as observed experimentally (dark dotted curve in Fig. 4a and other data in Fig. S10). At the chosen bias voltage ($V_s$ = -1.2 V), the contour of the detected photocurrent signal mostly follows the shape of the DOS distribution of the $Si_{\pi*}$ band because the energy range of the selected photons will match this delocalized surface state band. The contribution of resonant tunnel electrons at the energy of the STM tip Fermi level arises from the $Si_\pi$ band will remain a minor contribution as observed experimentaly[23].

At negative bias, additional significant source of photocurrent arises from the dissociation of free excitons in the conduction band of the silicon surface at the irradiation area[36]. Due to their longer lifetime than the free carriers at low temperature (> 10 μs), free excitons can move efficiently towards the STM tip apex location, while they are created further away from this position, over the irradiated area of the laser spot, to finally participate in the formation of the tunnel current via their dissociation[11,37,38]. Free excitons in crystalline silicon have a diffusion length of ~ 24 μm at 9 K. This effect allows a substantial number of excitons to reach the position where the tunnel electrons flow, at the STM tip apex, prior to their recombination[39]. The binding energy of free excitons in silicon worth 14.7 meV[40]. Excitons dissociation can be favored by the presence of an electrostatic field, vibrational excitations that are induced in the silicon lattice[41] as well as the presence of defects and interfaces[42,11]. All of these characteristics are present in the STM junction in our experimental conditions. In particular, since the band structure of the silicon has an indirect gap, it is required that the optical excitation transitions are associated with phonons for the formation of free excitons. The emergence of the necessary vibrational activation energies can arise from the inelastic processes prompted by the tunnel electrons flux in addition to the slight temperature elevation ($E_{thermal@16K}$ ~ 1.38 meV) induced by the laser irradiation as depicted in Figs S7 and S15.

Therefore, the photocurrent peak observed at 1030 nm ($E_{1030nm}$ = 1.20 eV) is specific to the silicon surface and may be attributed to the presence of free excitons created at the silicon surface that suddenly participate to the tunnel current via their dissociation, as they reach the vicinity of the tip apex. The photovoltage process that significantly grows when the incoming photon energy is higher than silicon surface band gap enhances this threshold effect[43]. The photocurrent peak energy at 1030 nm is slightly exceeding the silicon surface gap energy threshold ($E_{gap}$ = 1.12 eV) by $E_{1030nm} - E_{gap}$ = 80 meV[44]. This small amount of excess energy is required to overcome the free exciton binding energy ($E_b \sim$ 15 meV). The remaining energy (65 meV, calculated as 80 – 15 meV) may be attributed to the energy associated with surface photovoltage-induced band bending on the silicon surface. Previous studies have reported that the energy of photovoltage-induced band bending can range between 5 meV and 100 meV on different silicon surfaces, depending on several parameters such as laser power, wavelength, dopant concentration, and temperature[45,46].

When the STM tip is located above an Er adatom, the formation of additional photocurrent peaks arises from very localized events that involve trapped excitons in the vicinity of the Er (Fig. 5a). Trapped excitons emerge from free excitons that are located within defects[47,48]. In the present case, the defects in our silicon substrates are mostly composed of Er adatoms or As dopants. Some of the As dopants can be considered as shallow elements due to the relatively high dopant concentration in our samples. However, the lifetime of trapped excitons at As donors is relatively short ~ 183 ns[49], thereby limiting their contribution to the photocurrent signal from the silicon surface. The formation of trapped excitons at embedded Er atoms in silicon is rather well documented when related to the interstitial $Er^{3+}$ cationic state[50]. Their trapping energy worth ~150 meV[51] and their lifetime is about five time longer than free excitons (~ 50 μs). This characteristic trends to enhance multi excitons trapping efficiency at the erbium adatoms location[52].

As described in Table 1, the observed photocurrent peaks appear at specific resonant energies, as a signature of the Er:4f → 5d or Er:4f → 4f transitions, triggering the dissociation of trapped excitons. To further describe the interactions between the optically excited erbium adatoms ($Er^*$) states and the formation of additional tunnel electrons, one can consider the $Er^*$ as a local exciton that can interact with its close environment (Figs. 5a, 5b). The intrinsic lifetime and energies of the excited 5d or 4f levels in the Er adatoms will rule the relaxation rate of the $Er^*$ ($R_1$ in Fig. 5b) and thus trigger the formation of photocurrent in the STM junction at negative bias (Fig. 5a). In this scenario, the relaxation of $Er^*$ might underlie diverse processes: (i) Er* relaxation can induce Auger excitation of trapped excitons near the Er atom, promoting the formation of additional photo induced electrons in the tunnel current ($A_1$ and $D_2$ in Fig. 5b). (ii) The relaxation of Er* may also induce the recombination of trapped excitons via energy transfer (Re processes in Fig. 5b). But this process may not produce additional electrons and can thus be considered as a quenching process for the formation of the observed photocurrent peaks[53]. (iii) Additional trapped exciton can also be created via resonant energy transfer process from the relaxation of Er* to the silicon surface. These extra excitons has the potential to undergo direct separation, resulting in the generation of photocurrent[54]. (iv) Trapped excitons can form virtual trapped states within the Er* adatom[55]. The relaxation of Er* to this trapped state (R2 in Fig. 5b) via charge transfer will lead to the creation of a negative trion[56]. The formation and dissociation of such particles may also contribute to the creation of the observed photocurrent peaks[57].

It is important to notice that the thermalization time of the trapped excitons is long enough (~ 1 ns) for energy transfer processes induced from Er* relaxation[44]. However, the excitons formed within the delocalized $Si_\pi^*$ bands via $sp^2$ and $sp^3$ hybrid orbitals via electron scattering can lead to unfavorable exciton recombination rules as the trapped exciton dipole will be mostly oriented parallel to the surface ($Si_{up} - Si_{down}$ transitions) in the surrounding of the Er adatoms[44]. However, the light-phonon coupling energy of these transitions (~ 64 meV) related to the stretching vibrational mode energy in the $\Gamma X$ and $\overline{\Gamma}\overline{J}$ direction favor exciton propagation[58,59]. In this configuration, when associated with the exciton's vibrational coupling to $\overline{\Gamma}\overline{J}$ direction[58], some Auger excitation processes of trapped excitons and their dissociation may be favored [60,61]. Also, the relaxation of excited Er*, acting as a driving force, will favor the direct dissociation of trapped excitons when located at Er adatom at the surface, (processes $D_1$ and $D_3$ in Fig. 5b). Indeed, the formation of photoelectrons is shown to be activated by the excited Er* relaxation and favored by various experimental parameters such as: (i) the presence of an electrostatic field, (ii) the induced local vibrations via inelastic tunneling processes and (iii) the presence of the interface Si/Er/vacuum[11,42,62,63].

It is therefore interesting to explore the possible origin of the measured photocurrent peaks linewidth broadening. Here we will not consider lattice broadening effect since the Er adatoms are sufficiently decoupled from the silicon bulk lattice. In this context, the measured photocurrent peaks linewidths broadening are clearly different for each Er adatoms and worth, in average, $<\Delta v> = 31.6 \pm 6.3$ cm-1 for $Er^{BB}$ and $<\Delta v> = 15.8 \pm 5.1$ cm-1 for $Er^T$. These values can be compared with the calculated stretching mode of the Er adatom in the pedestal conformation where vibrational frequencies varies from 45 cm$^{-1}$ to 71 cm$^{-1}$. This reveals that the induced vibrations of the adatoms may be correlated to the broadening of the photocurrent peaks and, hence, to the formation of the photocurrent itself (see Fig. S13).

The formation of additional tunnel current may also occur via, various other processes such as singlet fission[64] involving one or several trapped excitons. A detailed study of these dissociation processes will require further investigations to elucidate the dynamics of the Er excited states and its relaxation in its environment such as, for example, the analysis of light emission from the Er adatoms. This is beyond the scope of this work and will be the object of future investigations.

**Conclusion**

Our experimental method can accurately describe the electronic structure of Er adatoms on the Si(100) surface at the atomic scale through precise localized photocurrent measurements. The observed photocurrent signal arises from a range of various photoinduced processes, including the formation of free carriers and excitons within the $Si_{\sigma*}$ and $Si_{\pi*}$ energy bands, which shape the photocurrent envelope. On top of this photocurrent envelope, additional photocurrent features are observed, involving trapped excitons at the photoexcited Er adatoms. This process enhances the detection of photocurrent peaks corresponding to specific wavelengths of the excited states of Er adatoms. With this method, we successfully differentiate between two specific photocurrent spectra corresponding to the two types of observed Er adatoms configurations. Each of the two Er adatom adsorption sites is modeled via the density functional theory method to describe their precise atomic interaction with the silicon surface and their electronic structure. Our DFT investigations on both adsorbed systems, which include spin-orbit coupling, indicate two different electronic configurations for $Er^T$:[Xe] $6s^1 5d^{0.9}(^1S_0)$ $4f^{11.4}(^3D_3)$ and $Er^{BB}$:[Xe]$6s^{0.9}5d^{0.8}(^1S_0)4f^{11.5}(^3S_1)$. This distinction is achieved by selectively exciting Er:4f → 5d and Er:4f → 4f transitions as the incident photons wavelengths vary from 800 nm to 1200 nm. Remarkably, the energy resolution of the detected photocurrent peaks (FWHM) can reach 5.8 x $10^{-4}$ eV (i.e. 4.68 cm$^{-1}$). The estimated photoelectron yield for the observed Er transitions is ~ 8 x$10^{-3}$ electron/photon. Our robust spectroscopic method can extend to any type of adatom or molecule on the silicon surface. Specifically, this technique could also be employed to address optically and very locally, any lanthanides adatoms to prepare specific quantum state at the atomic scale. Hence, the versatility of our spectroscopic technique lays the groundwork for the development of future devices that can be optically addressed and electronically probed simultaneously. Ultimately, 2D materials in which efficient free and trapped excitonic effects are conceivable could be a good candidate to explore similar phenomena using adsorbates with long-range surface interactions[11].

**Acknowledgements**

D.R. would like to thank the LABEX PALM for the *Investments for the Future* financial support ANR-10-LABX-0039-PALM as well as T. Chanelière and A. Louchet-Chauvet for valuable discussions. D.R. would also like to thank Thomas Ferhat, from NKT Photonics, for the loan of the supercontinuum laser and Juganta Roy for proof reading of the manuscript and appreciated discussions. DR will finally thank Profs. C. Ropers, B. Schroeder, M. Wenderoth and M. Weinelt for valued discussions. E.D. wishes to thank the Mesocentre de calcul of the Franche-comté University and the *Communauté d'Agglomeration du Pays de Montbeliard* (convention PMA-UFC). All the authors discussed the results and contributed to the manuscript writing. D.R. supervised the project.

**Data availability**

The data supporting the findings of this study are available within the main text of this article and its Supplementary Information. Further information about this study is available from the corresponding authors upon reasonable request. The source data are provided with this paper.

**Authors information**

Dr Eric Duverger, ORCID N° 0000-0002-7777-8561, eric.duverger@univ-fcomte.fr

Dr. Damien RIEDEL, ORCID N° 0000-0002-7656-5409, damien.riedel@universite-paris-saclay.fr

Corresponding author: Dr. Damien RIEDEL, web site: http://mnd-sciences.com

**Correspondence**

Any request should be addressed to the corresponding author of this article

**Authors contributions.**

D.R. mainly wrote the article, performed the experimental measurements and analysis of the data as well as numerical simulations with Gaussian 09 software. E.D. has done the DFT simulations using the SIESTA code of the $Er^T$ and $Er^{BB}$ configurations, relaxation of the atomic structure, PDOS on various atomic orbitals.

**Competing interests**

The authors declare no competing financial interests.

**Additional information.**

A Supplementary information document is available for this article.

**Figures Captions:**

**Figure 1| STM topography and dI/dV measurements on $Er^{BB}$ and $Er^T$ adatoms adsorbed on the bare Si(100) surface.** (a) Ball-and stick sketch of the STM junction showing an Er adatom adsorbed on

the Si(100)-2x1 surface when irradiated by a pulsed laser at different wavelengths while reading the photocurrent. (b) STM topography (147 x 108 Å², $V_s$ = -1.9 V) of the Si(100)-2x1 surface after the adsorption of Er atoms. The $Er^T$ and $Er^{BB}$ stand for Er atoms adsorbed on top and in between two silicon dimer rows, respectively. (c) and (d) Ball-and-stick representation of the two Er configurations $Er^{BB}$ and $Er^T$, respectively. (e) and (f) comparison of the two STM occupied ($V_s$ = -1.5 V, 15 pA) and unoccupied ($V_s$ = 1.5 V, 15 pA) topographies (31 x 31Å²), first row, with DFT simulated STM images, second row, at the same energies, below or above the Fermi level energy, for the $Er^{BB}$ and $Er^T$, respectively. (g) and (h) dI/dV differential conductance curves acquired on the $Er^{BB}$ and $Er^T$ (red curves), respectively, compared with the dI/dV spectrum acquired on the bare silicon surface (dark blue curves). A detailed description of the atomic slabs used for our DFT calculations is provided in Fig. S16.

**Figure 2| Analysis of the photocurrent induced on the bare Si(100)-2x1 surface.** (a) Comparison of the relative evolution of the photocurrent (arbitrary units) acquired on the bare Si(100) surface at positive bias, the laser intensity and the voltage pulse as a function of the laser wavelength and time. A 38 x 38 Å² STM topography inset recalls the measure spot location on the bare silicon surface. (b) Evolution of the photocurrent signal intensity acquired on the bare Si(100)-2x1 surface at positive bias as a function of the laser wavelength for four different tip height variations $\delta z$. The STM parameters are set to $V_s$ = +1.25 V, I = 7 pA, which determine the initial value of the tip height dz ($\delta z=0$) at which is recorded the first less intense photocurrent curve. (c) Evolution of the photocurrent value as a function of the tip height variation $\delta z$ for different wavelengths 910 nm, 1030 nm and 1064 nm. For 1030 nm and 1064 nm, only the intensity of the photocurrent peaks, from the base of the photocurrent envelope to its maximum, is considered. The right (left) horizontal axis is assigned to data at 1030 nm and 1064 nm (910 nm), see blue and black arrows for the respective intensity axis. The dotted red and green curves are numerical fitting according to the mathematical expressions (d) Evolution of the photocurrent as a function of the surface bias for the three considered wavelengths 910 nm, 1030 nm and 1064 nm for $\delta z$ = 1.4 Å. The error bars are represented by the size of the circles.

**Figure 3| Comparison of the photocurrent variations on the Er$^{BB}$ and Er$^{T}$ with their calculated structures.** (a) and (b) absolute values of photocurrent intensity curves acquired on the Er$^{BB}$ and Er$^{T}$ adatoms as a function of the laser wavelength. The dark red curve is the result of an averaged smoothing of the light gray curve (considering ten samples on both sides of each measured point). The wavelengths at which the most intense photocurrent peaks are detected, are indicated on top of the curves. The inset in (b) shows a detail of the three narrow photocurrent peaks within the 930 nm – 1000 nm range. An STM topographies (31 x 31 Å²) inset indicating where the measurements is done is recall on each panel. The light blue voltage curve timescale is recalled for clarity with $V_{exc.}$ = -1.2 V in both cases. (c) and (d) spin polarized representation of the Kohn-Sham orbital energies levels of the Er:5d and Er:4f orbitals extracted from the PDOS curves calculated after the DFT relaxation of each atomic slabs as presented in Figs. 1c and 1d, for the Er$^{BB}$ and Er$^{T}$ adatoms configurations, respectively. For each panel, the spin up and spin down energy levels are symbolized by short horizontal lines in dark gray, red, blue and green for the 5d$_{up/dn}$ and 4f$_{up/dn}$ orbitals, respectively. The minimum and maximum wavelengths for possible Er:4f $\rightarrow$ 5d transitions are indicated for comparison. The names of the occupied Er:4f orbitals energy within $E_f$ are indicated. On the right side of each panel (c) and (d), a simplified band diagram of the Si(100) surface is recalled (blue rectangles) as well as the dI/dV curves acquired at each Er adatom (red curves) for comparison.

**Table 1| Comparison between the calculated transitions with DFT + SOC and the photocurrent peaks wavelengths measured on Er$^{BB}$ and Er$^{T}$ adatoms.** List of the calculated optical Er:4f $\rightarrow$ 5d and Er:4f $\rightarrow$ 4f transitions wavelengths for the Er$^{BB}$ and Er$^{T}$ adatoms configurations and the experimental photocurrent peaks. The calculated wavelengths (dark blue column), wavenumbers and names of the orbitals involved, as well as the detailed spectroscopic changes (last column) in the transitions are given. Comparison with the measured photocurrent peaks wavelengths ascribed to the calculated one (brown column), their wavenumber as well as the wavenumber difference with the ensuing calculated transition ($|\Delta\nu| = \nu_{Calc.} - \nu_{Exp.}$). Other quantum numbers variations such as $\Delta m_l$, $\Delta s$ and $\Delta J$ are indicated as extracted from DFT simulations.

**Figure 4| Band diagrams describing the photoexcitation processes in the STM junction and the formation of photocurrent at positive and negative bias voltages.** (a) and (b) Energy (on scale) band diagrams of the induced optoelectronic processes in the STM junction at positive ($V_s = 1.2$ V) and negative ($V_s = -1.2$ V) bias voltage, respectively, when the STM tip probes the bare silicon surface (i.e. far from Er adatoms) when irradiated at various wavelengths. The representation of the energy distribution of the unoccupied $Si_{\pi*}$ and $Si_{\sigma*}$ bands and the occupied $Si_\pi$ and $Si_\sigma$ bands are indicated. In each panel, the photo-creation of hot electrons in the tungsten tip, free carriers and free excitons (yellow ellipsoid) in the silicon are indicated, as well as the main direction of the electrons tunneling that form the photocurrent signal. The dotted orange curves in (a) and (b) are the profiles of the corresponding measured photocurrent on the bare silicon surface. In (a) the electrons tunneling backwards to the STM tip are indicated for the transition at 1030 nm (see text).

**Figure 5| Direct and in reciprocal space band diagrams describing the photoexcitation processes on erbium and the formation of photocurrent via excitons dissociations.** (a) Energy (on scale) band diagrams of the induced optoelectronic processes in the STM junction at negative ($V_s = -1.2$ V) bias voltage when the STM tip probes an Er adatom when irradiated by the laser at various wavelengths. The representation of the energy distribution of the unoccupied $Si_{\pi*}$ and $Si_{\sigma*}$ bands and the occupied $Si_\pi$ and $Si_\sigma$ bands are indicated. The creation of free carriers and trapped excitons in the unoccupied silicon band is indicated at various wavelengths as well as the main direction of the electrons and holes movements that form the photocurrent signal at the Er adatom. The orange curve is a profile of a typical photocurrent spectrum acquired on an $Er^T$ adatom, for comparison. The green horizontal lines indicate the characteristic energies (not to scale) of the Er:4f and Er:5d orbitals of the Er adatom. (b) Simplified band diagram of the Si(100) sample along the Γ L, Γ X and X K reciprocal lattices directions showing possible interactions of the Er excited state with its surrounding bond excitons. For clarity, the energies of the erbium electronic states are not depicted to scale. The formation of excitons is combined with phonon ($k_{vib.}$) for momentum conservation due to the indirect silicon band gap. The dotted parabola at the trapped

excitons excited states symbolizes a virtual state (light brown dotted curve). The arrows $R_1$, $R_2$, $A_1$, Re, $D_1$, $D_2$ and $D_3$ stand for the various processes: Er relaxation, Er relaxation to exciton virtual state, Auger dissociation, exciton recombination and three exciton dissociations channels, respectively.

**Methods**

**STM imaging and sample/tip preparation**

All STM topographies, dI/dV spectroscopic curves and photocurrent measurements were performed using a low-temperature STM Beetle-Besocke pan-scan head from CREATEC running at 9K and equipped with lateral $CaF_2$ windows to ensure optical access from its side. The bias voltages are applied on the silicon sample while the STM tip is grounded, implying that occupied and unoccupied states are probed via negative or positive biases, respectively. The used silicon samples are very pure (10 x 5 mm²) Si(100) rectangles cut from a silicon wafer (ITME, As doped, $\rho$ = 5 mΩ.cm) and clamped between two molybdenum blocks to warrant macroscopic contact from the sample holder. The samples preparation is performed under a base pressure of 5 x $10^{-11}$ torr. It consists in slow (30 s) heating cycles (~ 950 °C) to remove the oxide layer followed by rapid (3 s) heating cycles up to 1150 °C combined with slow (1 minute) decreasing temperature plateaus to reconstruct the surface in a Si(100)-2x1 silicon dimers rows arrangement[65,24,23,19]. The number of reconstruction cycles is limited to 7 to avoid having significant diffusion of dopant atoms towards the surface that may induce long range interaction with the Er adatoms. The sample is then cooled down to ~ 12 K and placed in front of an effusion cell in which a crucible containing an Er rod is heated to ~ 1000 °C to reach the sublimation regime and control the evaporation of a small amount of Er atoms. The Er evaporation rate is initially calibrated in front of a quartz balance. The silicon sample is then exposed to the Er flux for ~ 2 s to obtain a coverage of about 0.07 ML. During the evaporation process, the base pressure in the UHV chamber does not exceed 7 x $10^{-11}$ torr and we estimate that the sample temperature change does not exceed 1K due to thermal radiation of the effusion cell. The sample is then transferred into the STM chamber.

In the STM chamber, once the STM thermal shields are closed, the pressure is lower than $2 \times 10^{-11}$ torr and at least ~ 1 order of magnitude lower inside the STM shields, preventing the bare silicon surface to be polluted by other elements for more than months. The metallic tips used for these experiments are made from pure tungsten wire (99.9 % W, 0.250 mm diameter from Sigma-Aldrich) and electrochemical etched in NaOH solution. The W etched tips are introduced in UHV and gently baked for 24 hours in a load-lock chamber at 120 °C. It is then placed in front of an electron gun to remove the $WO_3$ oxide residues that may cover the tip apex. For this work we have used two different samples and tips. The dI/dV spectroscopy measurements are performed via the use of two lock-in amplifiers, one is integrated inside the DSP of the STM controller. The second one is an external apparatus (SR830 from Standford Research Systems). The amplitude modulation and frequency of the first lock-in are fixed at 7 $mV_{pp}$ and 847 Hz, respectively. This frequency triggers the second lock-in device. Both results are acquired simultaneously and compared to avoid any possible spurious measurements.

**Laser and optical setup of the STM junction irradiation**

The used laser is a supercontinuum fiber laser (ERX20) from NKT Photonics having a repetition rate of 78 MHz and a tunable wavelength range varying from 400 nm to 2200 nm. The pulse duration is ~ 100 ps with a delay between two pulses of ~13 ns. The laser can deliver an average power ~ 6.5 mW with a peak power density of 13 $mW.nm^{-1}$ at the fiber pumping wavelength (1064 nm). The laser fiber is connected to a monochromator (model LLTF Contrast) from NKT Photonics that selects a specific wavelength from the incoming laser with a resolution of 2 nm, corresponding to an average energy resolution of ~ 2.8 meV in the explored range 800 - 1200 nm. The laser beam is introduced inside the STM chamber after three reflections on silver coated mirrors providing a good reflectivity for the considered wavelength range. In addition, these multiple reflections favor the selection of a linear and vertical polarization of the incoming laser beam (see Fig. S6). Therefore, the incoming light hitting the STM junction is mainly polarized in the direction of the STM tip, i.e. perpendicularly to the silicon surface. The laser beam is focused at the STM junction position via an achromatic lens ($BaF_2$, f = 16 mm from 400 nm to 1000 nm) fixed on the STM head. Hence, for an incoming laser beam of 2 or 3 mm, the laser beam waist diameter at the STM junction is ~ 50 µm. A great care has been given to keep the laser beam center at the middle of the STM junction.

**Temperature stabilization of the STM junction under irradiation**

The variation of the STM junction temperature during the laser irradiation can be roughly estimated by reading the temperature probe located at the STM head, nearby the STM junction (~ 2 cm) on the STM scanner (Si diode DT-670 from Lake Shore). Our measurements, indicate, that the STM junction temperature is stabilized after ~ 30 mins to a temperature of 16 K (see Fig. S7). In these conditions, the ensuing thermal energy $E_{th} = K_B.T$ worth 775 µeV at 9K and 1.38 meV at 16K. The STM junction can then be normally used for measurements, once stabilized, while the laser junction is irradiated continuously for several hours, providing that the cryogenic reservoirs are full. In our experiment, to prevent abrupt contact between the tip and the surface caused by their differing thermal expansions, the STM tip is initially retracted by approximately 100 nm before laser irradiation. After stabilizing the temperature of the STM junction, the STM tip can be repositioned in tunnel mode, allowing the scanning process to proceed normally. Once the STM junction stabilizes under irradiation, the laser beam remains continuously active throughout all measurements. During irradiation, the total STM tip expansion can be estimated, from previous work, to be in the range of 10 to 30 nm[66,67]. Any type of preliminary measurements that does not require to shine the STM junction with the laser (e.g. for dI/dV tunnel spectroscopy) are performed before any laser irradiation period. To provide a more accurate estimation of the temperature increase at the STM junction during spectra acquisition, we performed a calculation of the temperature variation in the silicon (surface) and tungsten (tip) regions exposed to the laser beam. This was done as the laser wavelength ranged from 650 nm to 1200 nm, following the thermal stabilization of the STM junction. These calculations show that the average increase of local temperature at both materials does not exceed 21 µK/nm. Therefore, with the feedback loop open during spectrum measurements, the stability of the STM junction remains highly robust across the entire range of tested wavelengths. This results in a maximum additional surface retraction of approximately 20 pm for silicon and 8 pm for the tungsten tip expansion during a single spectrum (see Fig S15 and note 10 for details). Our experimental findings also reveal that the photocurrent peaks observed on the Er adatoms cannot be attributed to minor variations in the STM tip-to-surface distance induced by the laser wavelength scanning (e.g., due to changes in laser power). This observation holds true as the acquisition process mirrors that of obtaining the photocurrent spectrum on the Si surface.

**Estimation of the laser intensity and photoelectron yield at the STM junction**

The optical intensity applied on the STM tip can be estimated via the calibration of the optical transmission of each optical devices used along the laser beam optical pathway for the considered wavelength range. This concerns the monochromator (and its fiber, T = 65 %), the three silver coated mirrors (each transmits 95% of the incoming light), the three $CaF_2$ windows installed at the vacuum chamber entry and at the two thermal shields (each transmits 92 % of the incoming light) and finally, the focalization lens made of $BaF_2$, which concentrate the light at the STM junction, transmits 90 % of the incoming light. Hence, the total optical pathway transmits 39 % of the incoming light. Here we deem that the optical transmission is almost flat for each optical device over the considered laser spectrum. The impact of the laser spot on the STM junction is decomposed into two main areas. The first area forms the main focal spot having a diameter of 50 μm shining the silicon surface and correspond to the projection of the circular laser beam onto the surface. The second area is an estimate of the section of the incoming laser beam that irradiate one side of the STM tip (see Fig. S8a for more details). The electromagnetic field that is confinement underneath the STM tip as the effect of the optical polarization of the tungsten tip and the silicon surface by the incoming laser light is estimated. Our calculation show that the electromagnetic field gain (enhancement) $G(\omega)$ located at the silicon surface underneath the tip apex indicates that for an STM tip having a radius of curvature R = 100 nm, the enhancement worth $G(\omega) = 1 + 8.85 \times 10^{-5} \sim 1$ and $G(\omega) = 1.078$ for R = 50 nm. Therefore, in these conditions, the electromagnetic field intensity that irradiates the STM junction is very similar to the one that is located underneath the STM tip, where the Er adatom is irradiated. Details of these calculations are provided in Figs. S11, S12 and supplementary notes N2 to N6.

**Recording of the photocurrent**

To record the photocurrent, the STM tip apex is firstly moved at a given position on the surface in constant current mode, after imaging the surface (i.e. either on a bare silicon area or on top of an Er adatom). Once this target position is reached, the feedback loop of the STM is opened. After a delay of 1s, the bias voltage is changed to the desired photocurrent reading parameter (i.e. usually ±1.2 V) and the STM tip is slightly retracted (or approached) from its initial tip-surface distance. This variation of voltage pulse combined with a variation of tip-surface height generates a trigger signal that is sent to the monochromator with a defined delay $\delta t_s$ to start the wavelength scanning. When the voltage pulse reaches 49s, the bias and the tip height are set to their previous scanning values. At the end of the spectrum reading sequence, the feedback loop is closed. For each spectrum, the photocurrent recording time lasts 50s (see Fig. S10). The STM tip is then moved, in constant current mode, by the STM software to a 'waiting' position before the next STM image or photocurrent measurement while the laser stays on at the waiting wavelength of 650 nm. To check that the tip height remains constant during the photocurrent recording process while the feedback loop is open for a duration of 50 s, a tunnel current trace is recorded with the laser irradiating the STM junction at a fixed wavelength. From our measurements, one can observe that the current trace is very stable during 48 s and worth, for the two given examples, +/- 4.06 ± 0.13 pA. Details of the photocurrent recording method is presented in Fig. S10 and note 5.

**Numerical simulations**

Numerical simulations of the two studied Er adatoms adsorption sites on the Si(100)-2x1 surface are carried out using the density functional theory (DFT) method integrated in the Spanish Initiative for Electronic Simulations with Thousands of Atoms (SIESTA) code with a relativistic pseudo potential, as usually applied for various lanthanide[68,69,70,71,72]. The norm-conserving relativistic pseudopotentials are generated using an input file made available to users by the SIESTA and ABINIT team[73,74,75]. Using the Troullier and Martins scheme integrated in the ATOM program, the ground state electronic configuration of the neutral Er is defined as [Xe]$4f^{12}6s^2$ while the $5d^0$ and $6p^0$ orbitals are initially empty. The calculations are conducted in two steps. In the first stage, a self-consistent field (SCF) convergence is achieved using a polarized double $\zeta$ basis set (DZP) and the spin-polarized option. The accuracy of this stage ensured reliable electronic structure calculations. In the second stage, spin-orbit coupling (SOC) calculations are performed based on the on-site approximation method, as described in references[76,77].

For the exchange correlation functional, we used the generalized gradient approximation (GGA) including the van der Waals interactions[78,79,80]. The mesh cutoff of 250 Ry with a grid of 4 x 4 x 2 k-points giving a total of 16 k-points in the irreducible Brillouin Zone is considered to calculate the total energies within a numerical precision of 1 meV. Geometry relaxation was performed by the conjugate-gradient method with the force convergence criterion of 0.02 eV/Å in the same slab volume (i.e., 2.32 x 3.09 x 2.5 nm$^3$) containing 337 atoms, including one As dopant atom, 239 Si, 96 H and 1 Er. The charge distributions of the Er adatoms and the surface is analyzed using the Bader charge analysis and a full decomposition of the DOS over all orbitals (i.e. total number of Sankey-type orbitals: 36). This is performed in the energy range -20.00 to 10.00 eV using a smearing energy of 0.2 eV, and 500 energy points in this range to generate projected density of states (PDOS)[81,82]. To compute the PDOS decomposition of the Er adatoms for any of the 36 spin-up and spin-down orbitals components, we firstly consult the orbital index generated after the self-consistent field SIESTA optimization. In this file, we find the list of all the generated quantum numbers n, l and m to describe the different orbitals in a spherical basis set. Then, using the *fmpdos* utility software included in the SIESTA package, we can retrieve the PDOS outputs, for the initially defined energy range, by introducing the ensuing quantum numbers of the desired orbital, integrated on the slab volume. The spin orbit coupling correction on the Er:4f and Er:5d orbitals is taken into account via the estimation of the excited states by using the PDOS curves of each orbital type. This approach allows us to determine several types of optical Er:4f $\rightarrow$ 4f or Er:4f $\rightarrow$ 5d transitions that induce excited Er$^*$ in a spherical basis. The Russell-Saunders term symbols, including the total angular momentum J, the magnetic quantum number $m_l$, the spin quantum number as well as the secondary total angular momentum $m_j$ are provided for each electronic state in a spherical basis and extracted from SIESTA results.

To reproduce the STM images, we computed the local density of states (LDOS) by using the block local DOS in the input siesta file. It permits to generate the LDOS file that is the DOS weighted by the amplitude of the corresponding wavefunctions, at different points in space. It is a function of energy and position. Using the *stm* utility software, we can generate STM images utilizing the Tersoff-Hamann approximation method to generate a map of the isodensity of state surface at the calculated LDOS energy window.

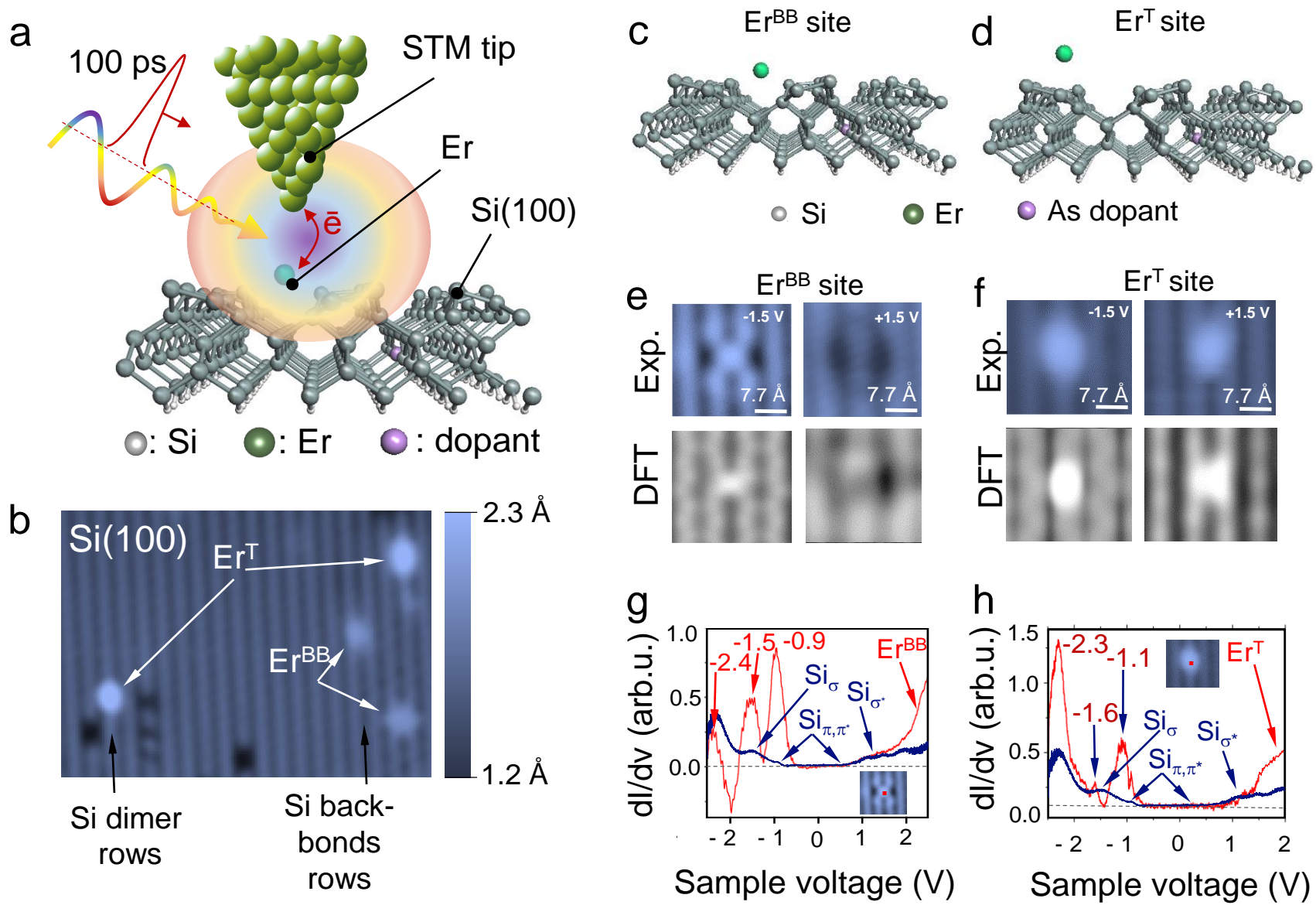

Duverger et al. Figure 1

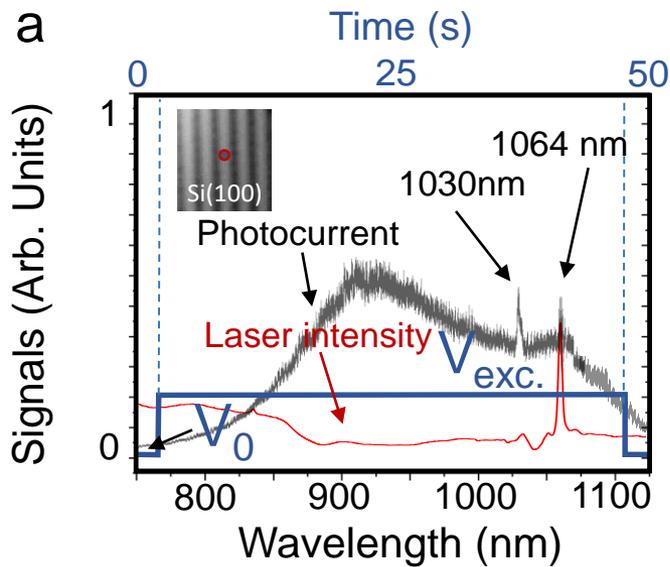
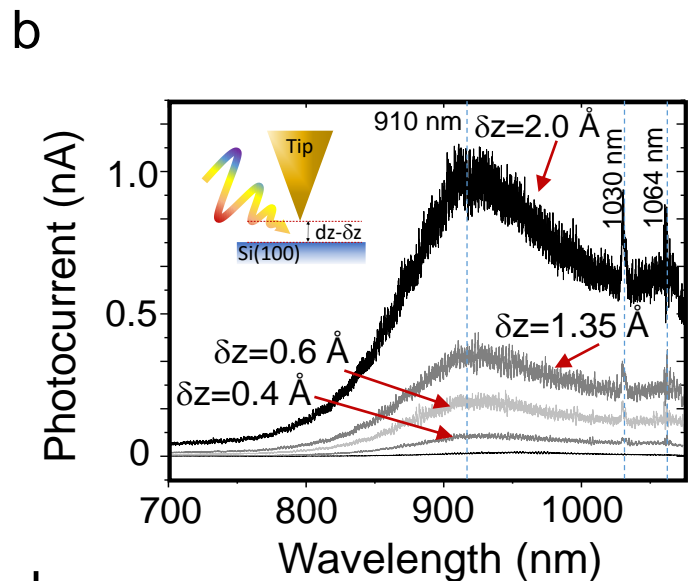
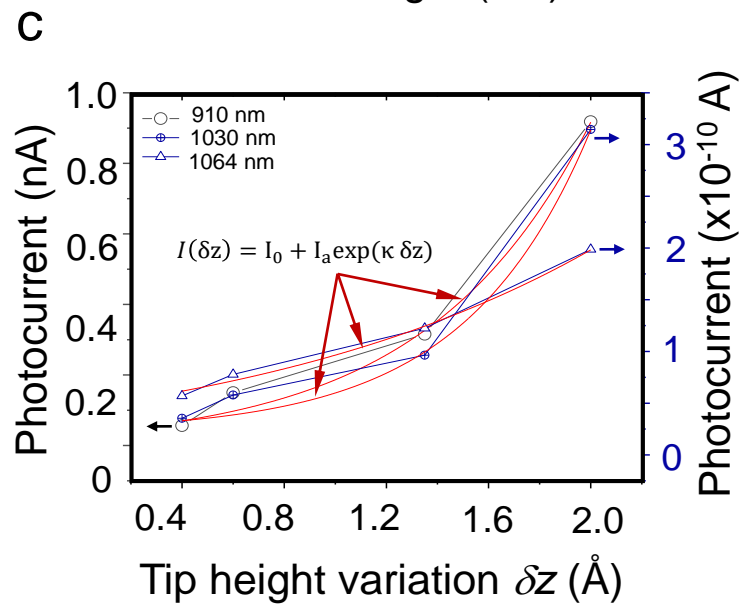
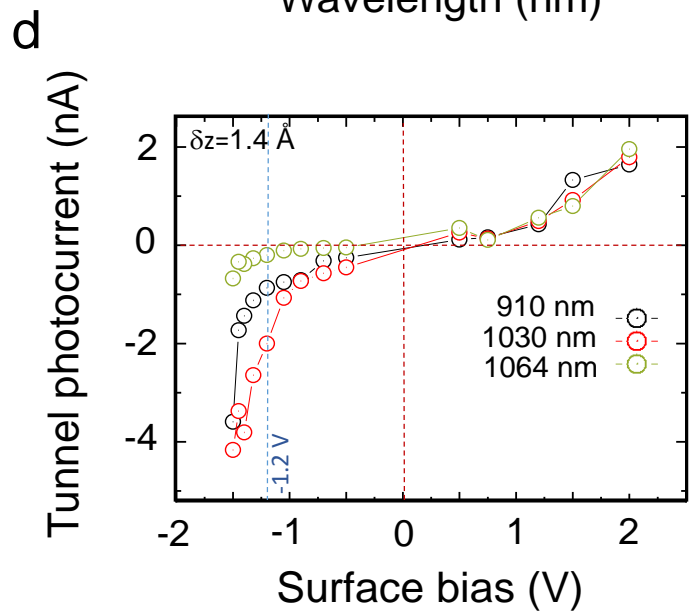

Duverger et al. Figure 2

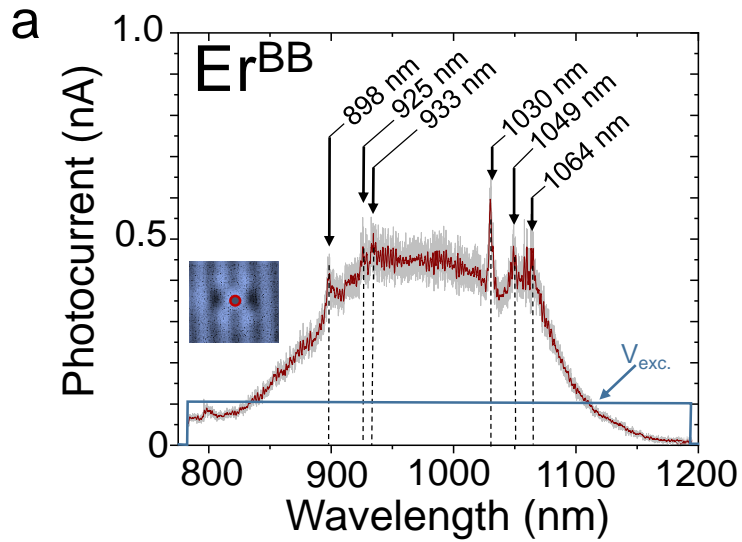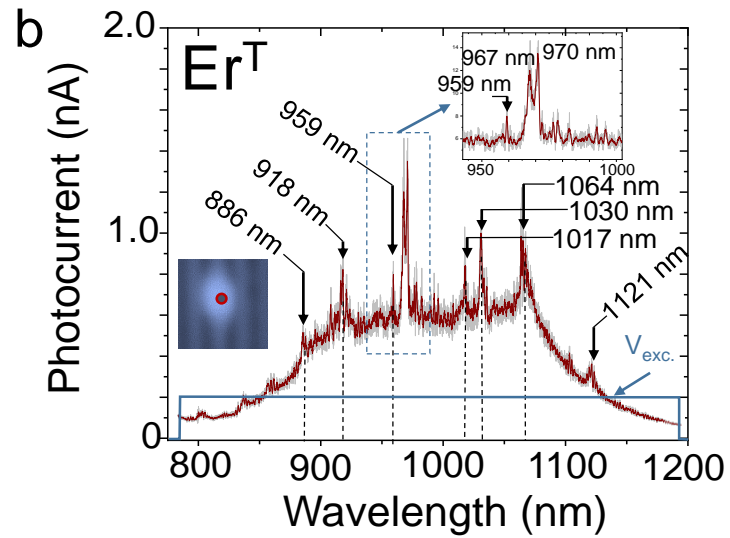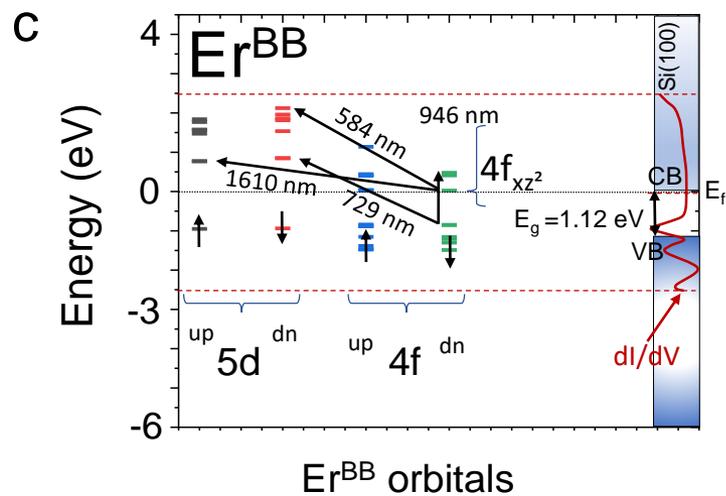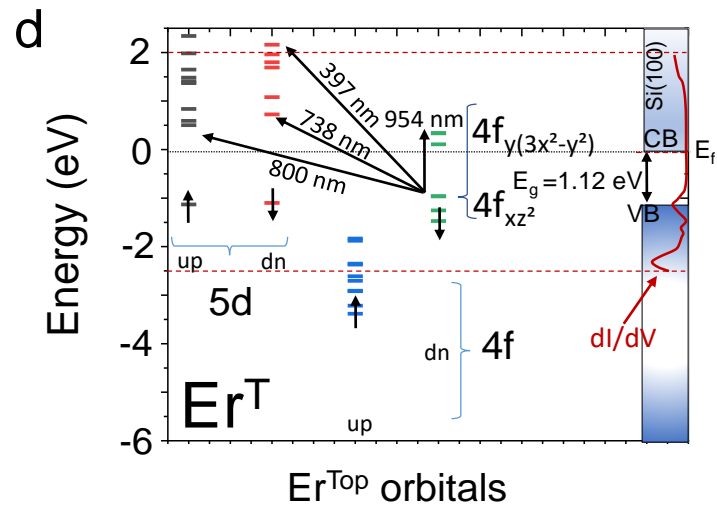

Duverger et al. Figure 3

| | λ Calc. (nm) | ν Calc. (cm$^{-1}$) | Transition orbitals | λ Exp. (nm) | ν Exp. (cm$^{-1}$) | \|Δ ν\| (cm$^{-1}$) | \|Δ ml\| | Δ s | Δ J | Transition type |
|---|---|---|---|---|---|---|---|---|---|---|
| Er$^{BB}$ | 890.42 | 11230.65 | 4f$_{\downarrow\ yz^2}$ → 5d$_{\uparrow\ z^2}$ | 898 | 11135.86 | 94.79 | 1 | 1 | 3/2 | 5d($^1S_0$) 4f ($^3S_1$) → 5d($^1S_0$) 4f($^2D_{5/2}$) |
| | 925.46 | 10805.44 | 4f$_{\uparrow\ z3}$ → 4f$_{\downarrow\ xz^2}$ | 925 | 10810.81 | 5.37 | 1 | 1 | 1 | 5d($^1S_0$) 4f ($^3S_1$) → 5d($^1S_0$) 4f($^3P_2$) |
| | 934.29 | 10703.31 | 4f$_{\downarrow\ z(x^2-y^2)}$ → 4f$_{\downarrow\ xz^2}$ | 933 | 10718.11 | 14.8 | 1 | 0 | 1 | 5d($^1S_0$) 4f ($^3S_1$) → 5d($^1S_0$) 4f($^3P_2$) |
| | 1047.31 | 9548.27 | 4f$_{\downarrow\ yz^2}$ → → 4f$_{\downarrow\ xyz}$ | 1049 | 9532.89 | 15.38 | 1 | 0 | 1 | 5d($^1S_0$) 4f ($^3S_1$)→ 5d ($^1S_0$) 4f ($^3P_2$) |
| Er$^T$ | 884.03 | 11441.25 | 4f$_{\downarrow\ xz^2}$ → 4f$_{\downarrow\ xz^2}$ | 886 | 11286.68 | 25.54 | 0 | 0 | 0 | 5d($^1S_0$) 4f($^3D_3$) → 5d($^1S_0$) 4f($^3D_3$) |
| | 911.09 | 10975.86 | 4f$_{\downarrow\ xyz}$ → 4f$_{\downarrow\ y(3x^2-y^2)}$ | 918 | 10893.25 | 82.61 | 1 | 0 | 1 | 5d($^1S_0$) 4f($^3D_3$) → 5d($^1S_0$) 4f($^3P_2$) |
| | 956.32 | 10456.75 | 4f$_{\downarrow\ x(x^2-3y^2)}$ → 5d$_{\downarrow x^2-y^2}$ | 959 | 10427.53 | 29.22 | 1 | 0 | 3/2 | 5d($^1S_0$) 4f($^3D_3$) → 5d($^1S_0$) 4f($^4F_{9/2}$) |
| | 961.52 | 10400.19 | 4f$_{\downarrow\ xz^2}$ → 4f$_{\downarrow\ xz^2}$ | 967 | 10341.26 | 58.93 | 0 | 0 | 1 | 5d($^1S_0$) 4f($^3D_3$) → 5d($^1S_0$) 4f($^5D_4$) |
| | 965.59 | 10356.36 | 4f$_{\downarrow\ y(3x^2-y^2)}$ → 4f$_{\downarrow\ y(3x^2-y^2)}$ | 970 | 10309.28 | 47.08 | 0 | 0 | 1 | 5d($^1S_0$) 4f($^3D_3$) → 5d($^1S_0$) 4f($^5D_4$) |
| | 1012.57 | 9875.86 | 4f$_{\downarrow\ xz^2}$ → 4f$_{\downarrow\ xz^2}$ | 1017 | 9839.84 | 36.02 | 0 | 0 | 0 | 5d($^1S_0$) 4f($^3D_3$) → 5d($^1S_0$) 4f($^3D_3$) |
| | 1111.59 | 8996.12 | 4f$_{\downarrow\ xz^2}$ → 4f$_{\downarrow\ z(x^2-y^2)}$ | 1121 | 8920.60 | 75.52 | 1 | 0 | 0 | 5d($^1S_0$) 4f($^3D_3$) → 5d($^1S_0$) 4f($^3D_3$) |

Duverger et al.   TABLE 1

Duverger et al. Figure 4

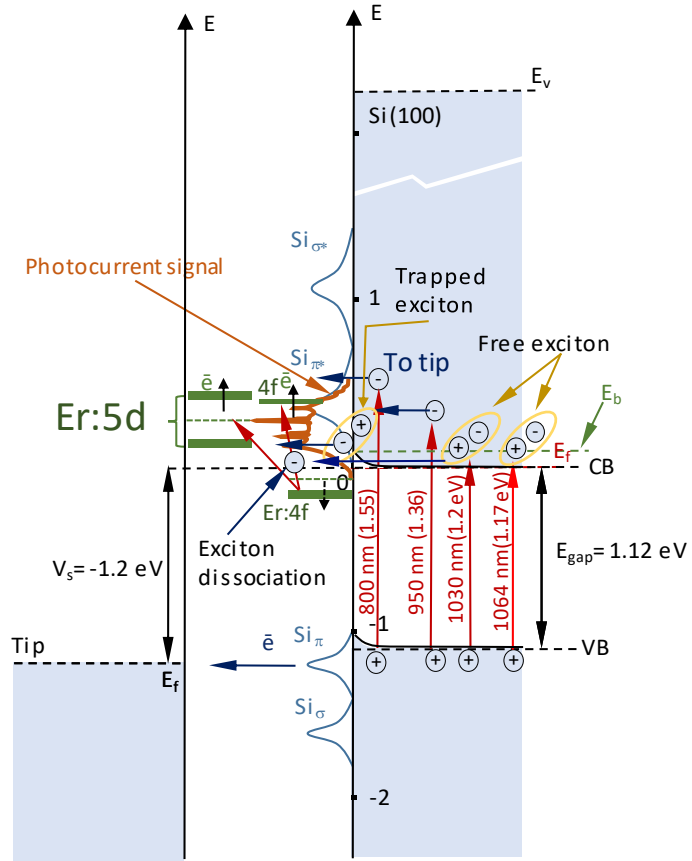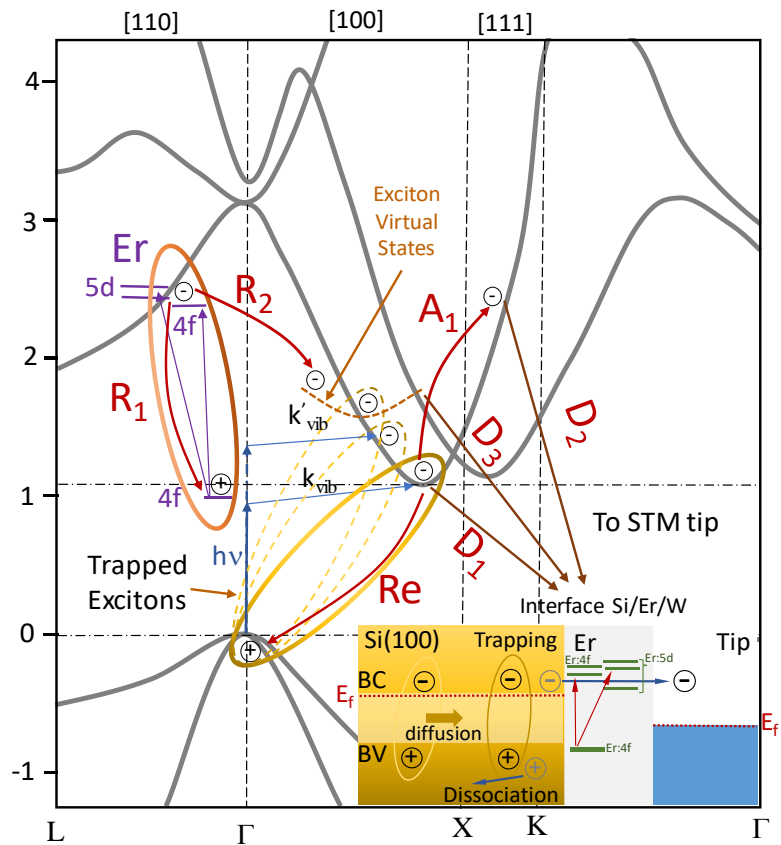

Duverger et al. Figure 5